\documentclass[preprint,3p,times,twocolumn]{elsarticle}

\usepackage{amssymb}
\usepackage{amsmath}
\usepackage{booktabs}
\usepackage{caption}
\usepackage{color}
\usepackage{subcaption}

\usepackage[hyphens]{url}
\usepackage{hyperref}
\hypersetup{breaklinks=true}

\journal{Nuclear Instruments and Methods in Physics Research A}

\begin{document}

\begin{frontmatter}



\title{Design and performance of an ionisation chamber for the measurement of low alpha-activities}


\author[ross]{ A. Hartmann}
\author[ross]{J. Hutsch\fnref{dead}}
\author[dres]{F. Kr\"uger}
\author[ross]{M. Sobiella}
\author[dres]{H. Wilsenach\corref{cor1} } 
\author[dres]{K. Zuber}

\cortext[cor1]{Corresponding author: \\heinrich.wilsenach@mailbox.tu-dresden.de}
\fntext[dead]{Deceased.}

\address[dres]{Institut f\"ur Kern- und Teilchenphysik, Technische  Universit\"at Dresden,  01069 Dresden, Germany}
\address[ross]{Helmholtz-Zentrum Dresden-Rossendorf (HZDR), 01328 Dresden, Germany}

\begin{abstract}

A new ionisation chamber for alpha-spectroscopy has been built from radio-pure materials for the purpose of investigating long lived alpha-decays. The measurement makes use of pulse shape analysis to discriminate between signal and background events. The design and performance of the chamber is described in this paper. A background rate of ($ 10.9 \pm 0.6 $) counts per day in the energy region of 1~MeV to 9~MeV was achieved with a run period of 30.8~days. The background is dominantly produced by radon daughters. 

\end{abstract}

\begin{keyword}

Frisch grid; Ionization chamber; alpha-decay



\end{keyword}

\end{frontmatter}


\section{Introduction}
\label{sec:int}


The study of alpha decay has been important for the understanding of nuclei and their properties for more than a century. Currently these studies still have impact in various areas of nuclear physics, providing information which is valuable and often not accessible otherwise. \\

The preformation of alpha-particles within nuclei has been a research topic since the beginning of nuclear physics and currently is in a phase of revival. The nuclides $^{12}$C and $^{16}$O are considered to have alpha particles as constituents of their underlying structure. This issue is intensively investigated world-wide (see  \cite{bij14}, \cite{Hoyle}).\\

The relation between the half-life and the Q-value has been established long ago as the Geiger-Nuttall law \cite{gei11}. Although the Geiger-Nuttall law is generally accepted, there is some fine structure involved which depends very sensitively on nuclear structure \cite{qi14}. Furthermore, empirical relations were established for a description beyond the Geiger-Nuttall law \cite{den09, den09a}. Studies of alpha-decays into excited states also provide useful information for nuclear structure studies \cite{wan09,wan10}. They might shed light on alpha-preformation and deformation in nuclei, especially in the vicinity of closed neutron shells \cite{cob12}. Exploring these level schemes would help to distinguish whether isotopes show deformation or have spherical cores with a preformed surface alpha. Furthermore, shape coexistence and intruder states can be explored by looking at the excited $0_2^+$ state \cite{wau94,del95}. Almost all of these studies are performed around the N=126 closed shell \cite{iac82,dal83} which is a reasonably well understood region. Fewer
studies have been performed around the N=82 closed shell \cite{iac82,iac85}, which is much less understood. The lack of quality data prohibits a more accurate study.

The half-life region between 10$^{6-10}$~yrs is very interesting as several nuclides in this range are used as geochronometers and cosmochronometers and the limitations on dating precision is sometimes restricted by the accuracy in the known half-life \cite{beg01,mar14}. Last but not least, another large area of alpha spectroscopy is the exploration of superheavy elements in the transuranium region and the actinide region itself.  \\
A completely different field of research, where alpha decays matter, is low background physics. Searches for neutrino-less double-beta decay, dark matter or neutrinos in general suffer from background of the natural U, Th radioactive decay chains. All materials used in the experiments have to be screened for radioactive contaminations, currently levels below 1 mBq/kg are required. Alpha spectroscopy can be a useful tool to identify and determine the activity in these materials.\\
To summarise, a variety of interesting topics can be explored with high precision low background alpha spectroscopy, which is the main purpose of the chamber described in this paper.

\section{Experimental Setup}
\label{sec:expset}

The detection principle of the chamber is based on ionisation. When an alpha particle travels through a medium, it loses most of its energy by ionising the atoms in this medium. A gaseous argon mixture (90\% Ar, 10\% CH$_{4}$ known as P10) is used as the ionisation medium. An electric field is placed across the chamber to separate and collect the electrons and the ions. The full collection time of the signal inside the chamber depends on the mobility of the ions and electrons inside of the gas. There is also a dependence on the position of the event. This has two major disadvantages, namely, the signal takes a long time to be fully collected ($\sim300~\mu$s) and the energy is smeared out due to the different ionisation paths caused by the angle of the event.

For this reason a grid is positioned between the anode and the cathode inside the chamber. The grid is placed on an intermediate voltage. This design is referred to as the Frisch-Grid method \cite{Frisch}. The use of the grid in the chamber splits the chamber volume into two sections. The interaction region (cathode to grid) and the collection region (grid to anode). 

The particle track is fully ionised inside the volume of the interaction region. Only the electrons then travel through the collection region. Since the signal only depends on the movement of the electron through the collection region, the signal collection time is faster ($\sim 1~\mu$s). All of the electrons now travel across the same distance before being collected, this removes the angular dependence of the energy detected in the chamber. A side effect of this method is that the ratio of the signal pulse in the collection region and that of the interaction region gives the relative position of the event. 

 As mentioned previously, the signal induced on the anode and the grid is due to the drifting of the ions (that are produced in the initial interaction process) caused by the electric field, and can be described by the Shockley-Ramo theorem \cite{Shockley}\cite{Ramo}. The chamber has been specially designed so that there is little interference from the charge carriers in the interaction volume on the anode. The ability of the chamber to shield the anode from the interaction region is commonly referred to as the grid inefficiency (GI). This was determined using the prescription given by A. G\"{o}\"{o}k in \cite{GI}, the uncertainty on the GI is purely statistical. The design parameters and grid inefficiency are given in Table~\ref{tab:chamber2}. 

\begin{table}
\begin{center}
\begin{tabular}{ll}  
   \toprule
  Parameter    & Value \\
   \midrule
   Chamber diameter [cm] &    30  \\
   Grid wire radius [mm]      & 0.075        \\
   Grid separation distance [mm]     &    2.0        \\
   Distance from grid to anode [mm]       & 35      \\
   Distance from grid to cathode [cm]       &   10.0   \\
   GI [\%]& 2.429(\textit{7})          \\
   $\frac{V_{anode}-V_{grid}}{V_{Grid}-V_{Cathode}}$ & 0.446 \\
   \bottomrule
  \end{tabular}
  \label{tab:chamber2}
  \caption{ The design parameters and measured grid inefficiency of the chamber. These parameters were chosen to maximise the energy resolution and minimise the grid inefficiency (GI).}\label{tab:chamber2}
 \end{center}
\end{table}

\begin{figure}
\centering
\includegraphics[width=\linewidth]{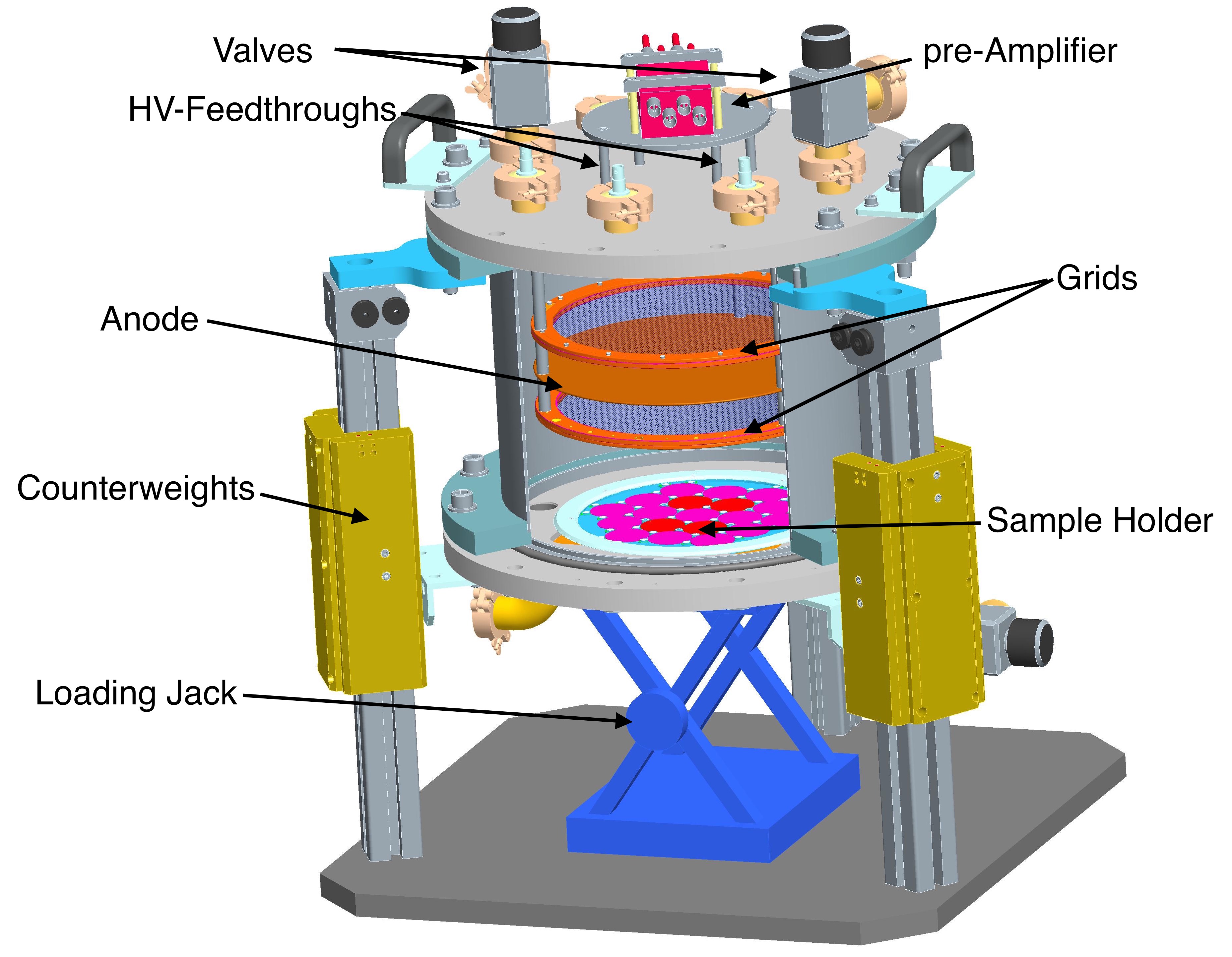}
\caption{Schematic drawing of the chamber. The grids and anode can be seen in the centre of the chamber. The first part of the signal manipulation can be seen on the top of the chamber, starting with the preamplifier. The valves are connected to the P10 canister for flushing. The jack and the counterweights are designed to allow for simple sample loading. The sample holder is interchangeable to fit different sample types and geometries.}
\label{fig:scem}
\end{figure}

The chamber was constructed using radio-pure materials. The main body of the chamber was designed using 
the largest electropolished stainless steel vessel guaranteeing a long term vacuum. The grid is constructed from a beryllium copper alloy, and the anode is made out of copper. The holders are made from radio-pure plastic. The holder material is shaped through laser cutting to keep the samples in position. The bottom flange of the chamber is lowered for easy sample loading. A schematic drawing of the chamber can be seen in Figure~\ref{fig:scem}. 

The chamber consists of an upper and lower section. The upper section acts as a general veto against noise and cosmic ray muons. There is a second design of this type of chamber which positions the sample in the centre of the two chambers to give 4$\pi$ coverage of the emission from a sample \cite{AlAdili2012103}, which is required for the application
of fission fragment spectroscopy. The design of the chamber described in this paper sacrifices $2\pi$ coverage for the ability to remove noise and ensure feasible long term low background running capabilities.

The chamber operates at three voltages. The anode is set at the highest voltage of 1200~V, and the grids in both chambers are held at intermediate voltages of 800~V each.  These voltages have been optimised to obtain the smallest energy resolution. The voltage also gives the most stable electron drift mobility, with respect to fluctuation in pressure due to the changes in temperature. 

The signal from the anode and the two grids are amplified through a CAEN A1422 Low Noise Preamplifier. The signal is then digitised with a CAEN N6724A sampling analog to digital converter (FADC). The FADC is set to trigger on the signal from the anode. The trigger level is determined to minimise the noise pulses caused by micro-phonics and electric noise. When the FADC is triggered, 20~$\mu$s sweeps from the anode and two grids are recorded. The FADC also records the time of each event. The wave forms from each channel as well as the temperature and pressure are all recorded in ROOT \cite{root} files, for easy data analysis. This does however limit the viable detection rate to less than about 10 Hz but upgrades for a faster readout
are possible. However, this is not crucial given the chambers main application is very low count rates. 


At the start of each run the chamber is flushed with P10 and afterwards sealed for the duration of the measurement. The chamber performs in a stable manner with the exception of a loss in detected energy as a function of time due to the degradation of the P10. For the future a continuous gas flow is planned to avoid this degradation. With multiple flushing of the chamber, this effect has been reduced to less than 2~keV per day. Currently this limits the total run time to about 40~days. The degradation in the ionisation of the gas has been linked to oxygen leaking into the chamber. An observed side effect of this (apart from the loss in ionisation as a function of time) is the decreasing in the signal collection time as a function of run time. The chamber is also run at a small over pressure to reduce this effect, $\sim$1040~mBar is the usual run pressure. This gives an E/p of ($0.105\pm0.004$)~V/(cm$\cdot$Torr). 


\section{Calibration and Performance}
\label{sec:cal}

\subsection{Energy Calibration}

\begin{figure}
\centering
\includegraphics[width=\linewidth]{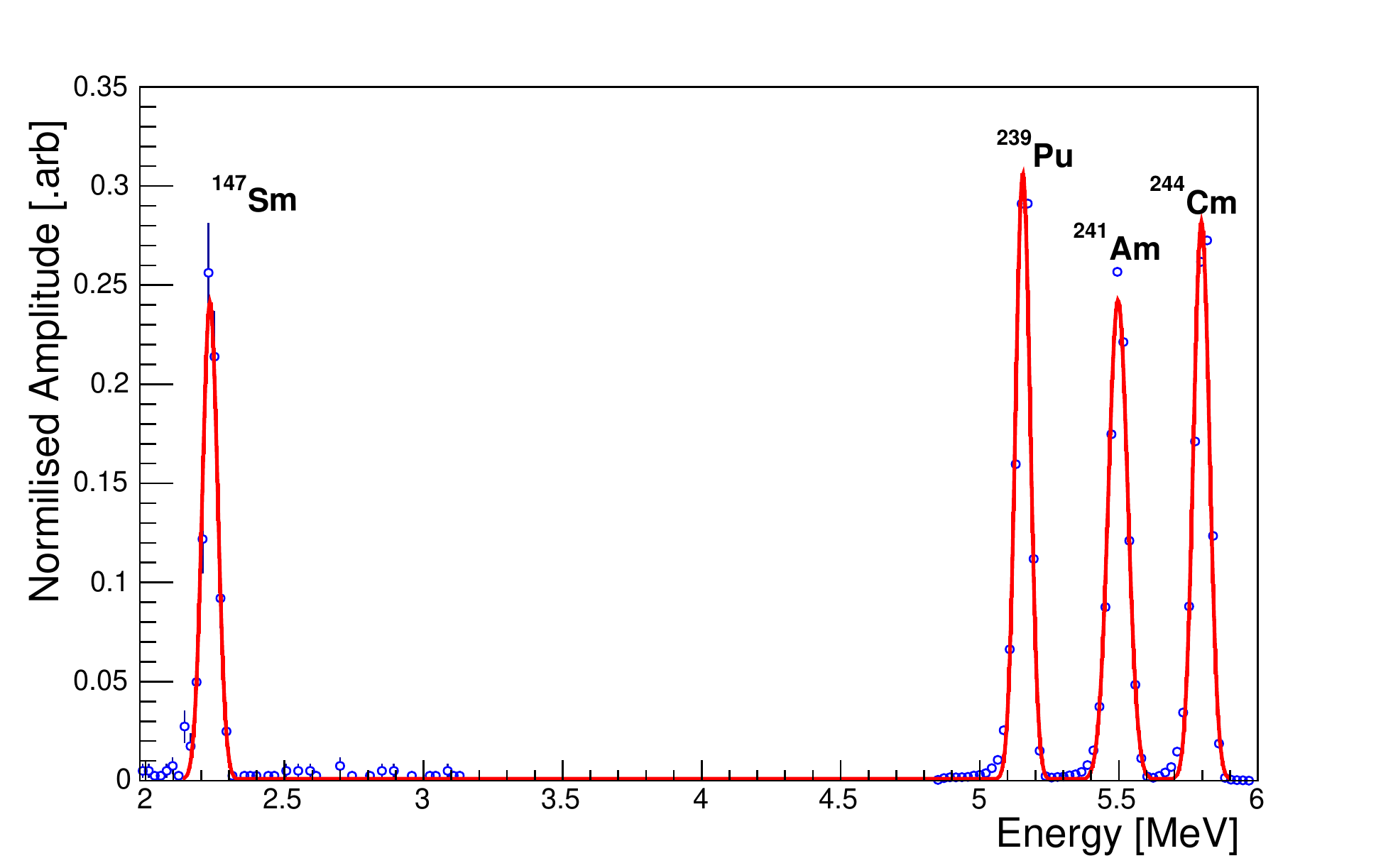}
\caption{The graph shows a calibration spectrum of $^{239}$Pu, $^{241}$Am, $^{244}$Cm and $^{147}$Sm. The area of each peak has been normalised to 1 for comparison reasons. The errors have been adjusted to take this into account.}
\label{fig:cal}
\end{figure}

The energy calibration was performed with two calibration sources. The higher energy calibration was performed using an alpha standard that contained three mixed nuclides, $^{239}$Pu (5156.59(\textit{14})~MeV \cite{a239}), $^{241}$Am (5.48556(\textit{12})~MeV \cite{a237}) and $^{244}$Cm ( 5.80477(\textit{5})~MeV \cite{a240}). The source was collimated due to its relatively high activity, this also helped to reduce the effect of energy loss inside the sample. The calibration at lower energy was performed using two $^{147}$Sm (2.248(\textit{1})~MeV \cite{a143}) samples with thicknesses of 40.20~nm and 31.40~nm respectively. The combined normalised spectrum is shown in Figure~\ref{fig:cal}.

\begin{figure}
\centering
\includegraphics[width=\linewidth]{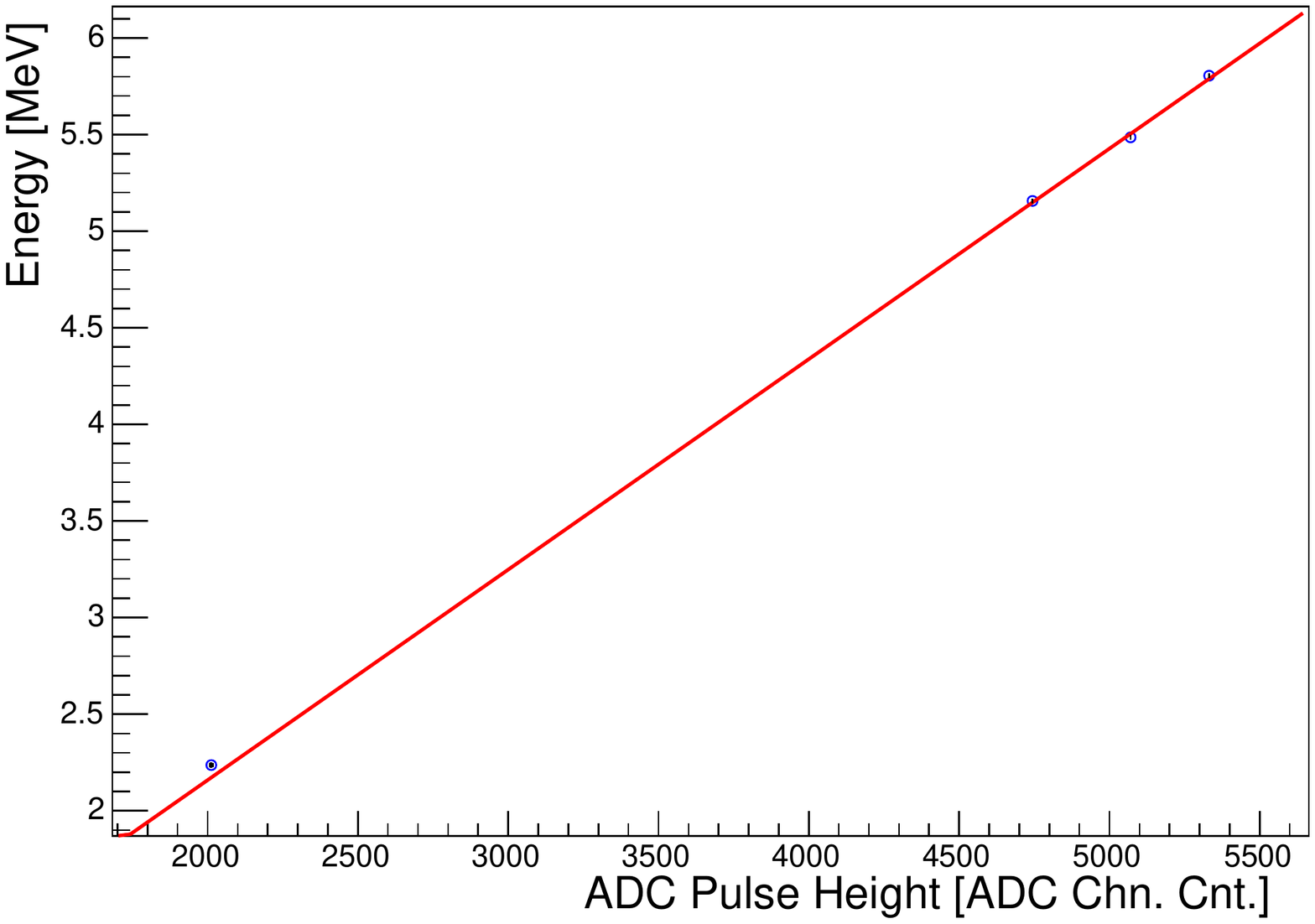} 
\caption{Energy calibration graph showing the energy as a function of FADC pulse height. The calibration has been fitted with a first order polynomial function. (Statistical error bars are present, due to the high statistics of the measurement they appear very small. The deviation is systematic in nature.)}
\label{fig:cal_graph}
\end{figure}

The energy as a function of pulse height is fitted with a first order polynomial. This is shown in Figure~\ref{fig:cal_graph}.
\begin{equation}
	E [MeV] = m \cdot Chan - c \;,
\end{equation}
where $m = 1.08961 \pm 0.00001$~keV/channel and $c = 20.86 \pm 0.07$~keV.

\subsection{Energy Response}

The resolution of the chamber is obtained with the same method as the calibration. Multiple gaussian distributions are fitted to each of the higher energy peaks. This is due to the fact that the decay of the high energy peaks are not directly to the ground state, but to multiple excited states. The energy resolution of the chamber is not good enough to distinguish the individual peaks of each isotope. The fit using multiple gaussian distributions only has the overall normalisation and the sigma of each peak as free parameters.
\begin{figure}
\centering
\includegraphics[width=\linewidth]{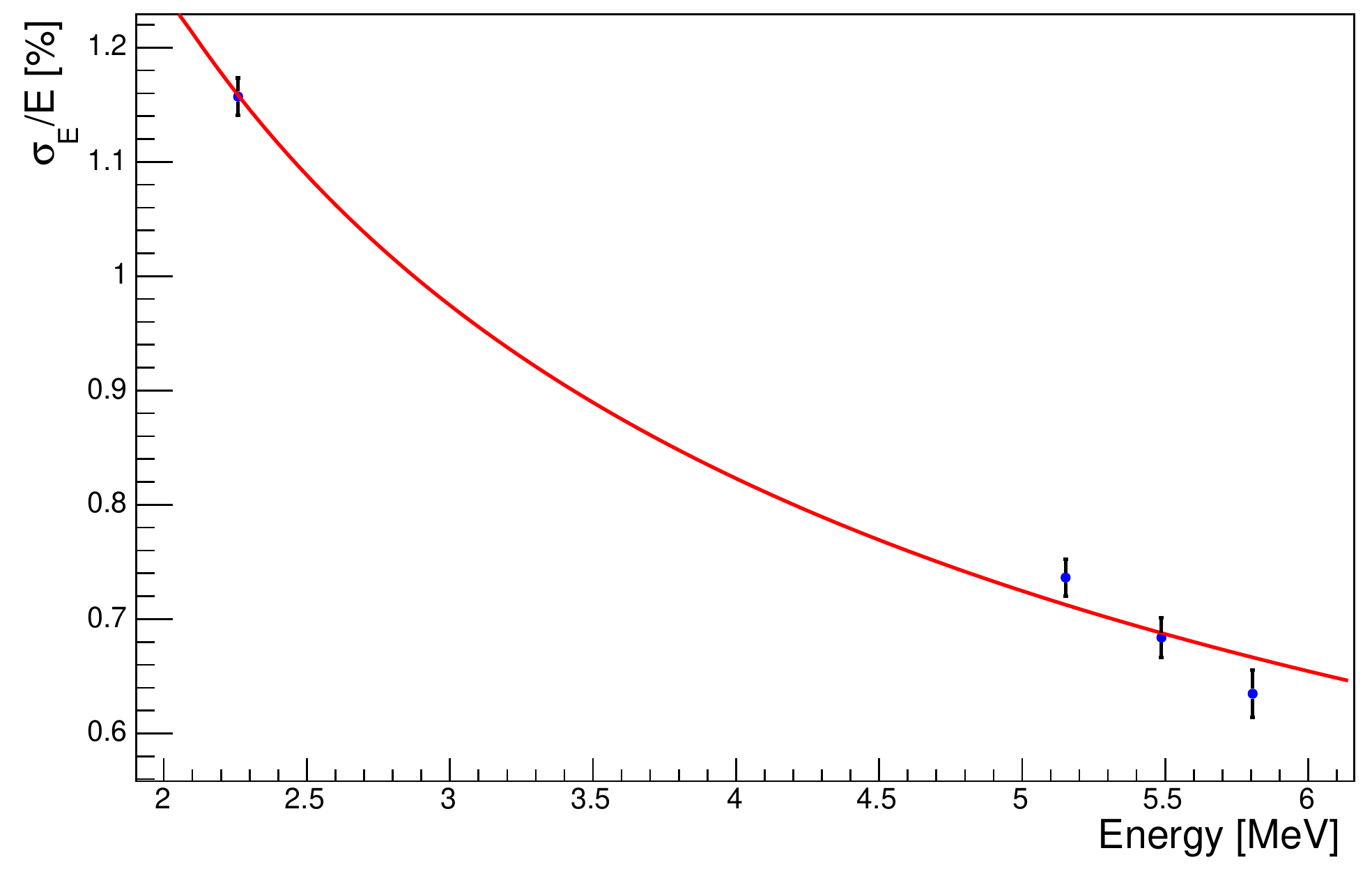}
\caption{The energy response as a function of energy from the decays of $^{239}$Pu, $^{241}$Am, $^{244}$Cm and $^{147}$Sm. The graph is fitted with the most general energy response function given in equation~\ref{eq:res}. The uncertainties are dominated by statistical variation. }
\label{fig:cal_res}
\end{figure}


It is assumed that the spread in energy resolution occurs because of three phenomena. The first is due to Poissonian variation of the current induced being proportional to the number of electrons collected ($\sigma_{\text{pois}}\propto \sqrt{E}$). The electrical response of the chamber is linearly dependent on the size of the pulses ($\sigma_{\text{elec}} \propto E$). The last variation comes from the energy independent smearing caused by noise ($\sigma_{\text{noise}} \propto$ const.). If all of these effects are assumed to be gaussian, the equation describing the energy response is then obtained by adding the variances of each effect in quadrature. Equation~\ref{eq:res} is used to fit the peaks in the calibration spectrum, the fit to data is shown in Figure~\ref{fig:cal_res}.

\begin{equation}\label{eq:res}
	\frac{\sigma}{E} = \sqrt{a^2+\frac{b^2}{E}+\frac{c^2}{E^2} } \;,	
\end{equation}

where $a$ is the factor of the electronic response, $b$ is the factor related to the Poissonian variation and $c$ is the constant noise term. The advantage of this parameterisation is that the factors are forced to be positive. This is usually simplified if one of the factors is much smaller compared to the others. 

It should be noted that the data used for this fit was obtained through the use of a collimated source to lower the rate and to reduce the smearing effect due to energy loss inside of the source. The angle of the alpha particle interaction was also constricted within the data to further reduce the self absorption. The detected energy loss as a function of time has also been corrected for in the data using the $^{239}$Pu peak as a reference.

\subsection{Run Stability}

\begin{figure}
\centering
\includegraphics[width=\linewidth]{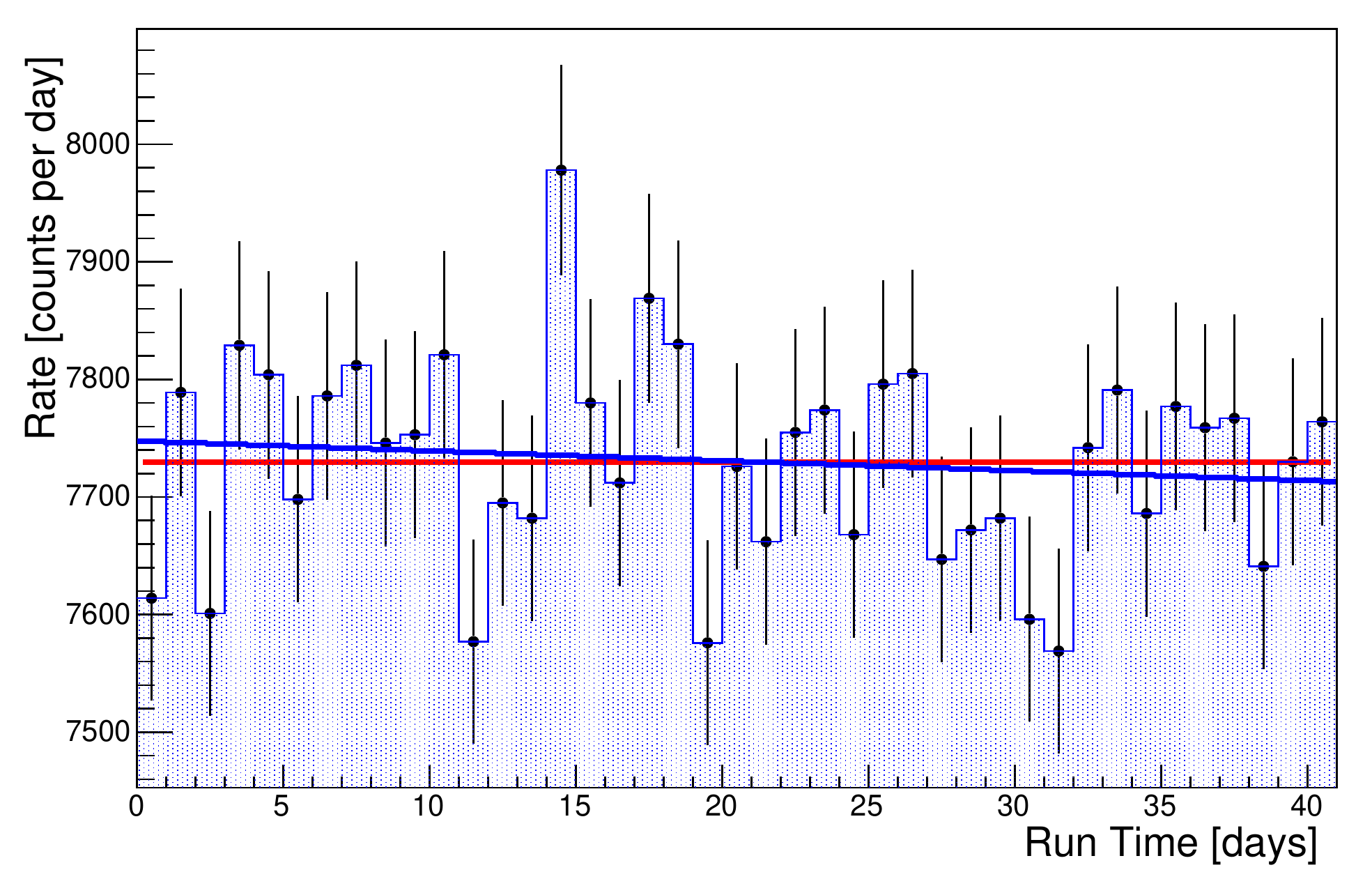}
\caption{Count rate of a collimated $^{241}$Am source as a function of 41~days of run time. The rate is fitted with a constant (red line) and a first order polynomial (blue line), this was done to show that there was no significant changes in the rate as a function of run time. }
 \label{fig:cal_rate}
\end{figure}

The chambers stability has been measured over a period of 41 days with a collimated $^{241}$Am source. The result of this measurement is shown in figure~\ref{fig:cal_rate}. (Only statistical errors were considered here.) The histogram was fitted with a constant and a first order polynomial. The value obtained for the rate using a constant is $7730\pm14$ counts per day (c.p.d), $\chi^2$/n.d.f. = 40.04/40. The first order polynomial does show a slight decrease in the region of ($0.8 \pm 1.1$) counts per day, and has a similar $\chi^2$/n.d.f. = 39.52/39 . There is no comparison for this value as the source that was used had no accurate activity determination. The rate would also differ from the overall activity due to the collimation of the source. Though there is a small decrease in count rate, the chamber was deemed stable for this chosen period of run time.

\subsection{Position Calibration}

As mentioned in Section~\ref{sec:expset}, the ratio of the anode pulse and grid pulse gives the relative position of the interaction. The relation is as follows,

\begin{equation}\label{eq:pos}
	x\cdot\cos(\theta) = X_0 \cdot \frac{|P_{Grid}|}{|P_{Anode}|} \;,
\end{equation}

where $P_{Grid}$ is the pulse height induced on the grid (the lower chamber grid), and $P_{Anode}$ is the pulse height induced on the anode. $X_0$ is the size of the interaction region, 10.0 cm. $x \cos(\theta)$ is the centre of charge of the ionised electrons multiplied with the angle at which the alpha particle leaves the sample. An angle of zero corresponds to an alpha particle that is going straight towards the anode. This factor can also be thought of as the projection of the centre of ionisation onto the x-axis.  

\begin{figure}
	\includegraphics[width=\linewidth]{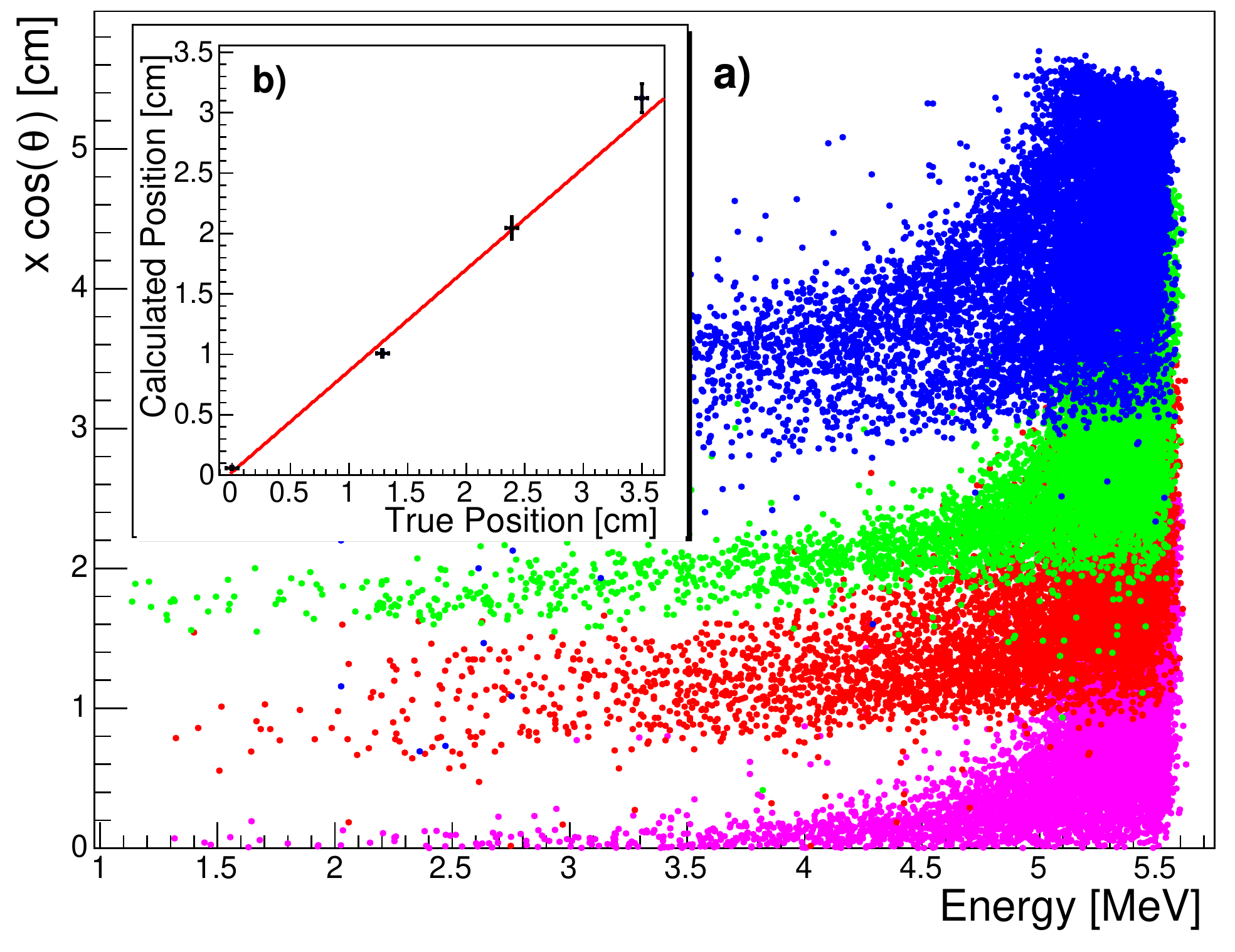}
	\caption{Graphs showing calibration of the vertical position inside of the chamber as function of energy. The sample was placed at 0~cm (pink) ,1.3~cm (red) ,2.4~cm (green) and 3.5~cm (blue) above the floor of the chamber. The inlet shows the position of source vs. the reconstructed position using equation~\ref{eq:pos}. }
	\label{fig:cal_pos}
\end{figure}

To calibrate the position resolution of the chamber, a weak $^{241}$Am source was positioned at different heights inside the chamber. The minimum and maximum positions were then calculated for the individual measurements. This is shown in Figure~\ref{fig:cal_pos}. The minimum value of $x \cos(\theta)$ corresponds to the surface of the source. The values for $x \cos(\theta)$ were entered into a histogram and the lower edge was fitted with a Gaussian. The value at half the maximum was used as the position reconstruction shown in Figure~\ref{fig:cal_pos} subfigure b. It can be seen that the position reconstruction is in good agreement with the measured position of the source. 

\section{Analysis, Cuts and Efficiency}\label{sec:cut}
\label{sec:ann}

The data acquisition (DAQ) is designed to record every pulse shape, this was done as only low rates are expected. Pulse shape analysis also made the development of a set of cuts based on physical phenomena possible. These cuts were developed to increase sensitivity to alpha events. The data cuts make use of the drift time of electrons in P10, as well as the position of the event. These physical phenomena have been used to impose four types of cuts onto the data. The cuts were experimentally developed using a $(3.457 \pm 0.076)$~Bq $^{241}$Am source running for 17.7~days. Data from background runs have also been used. The efficiency was determined using the same source.

\subsection{Noise and Cosmic Ray Muons Cut} \label{sec:noi}

As mentioned in section~\ref{sec:expset}, the chamber is constructed such that there are two segments, one on top of the other. Each segment is a fully functional detector. The same DAQ is applied to both chambers. The main use of this is to veto background components like noise. The main characteristic of the noise pulses are that they occur equally on each of the signal lines. A simple cut is set on this fact. If there is a signal in the top chamber as well as the bottom chamber, then the event is marked as noise. This cut also acts as a coincidence cut, as the cut would veto events of any signals caused in both chambers. This is also the typical signature of cosmic ray muons.  
\subsection{Trigger Order Cut}

If a signal originates from the bottom of the lower chamber, the charge cloud caused by the ionisation has to first move through the interaction region and then the detection region. This gives a well defined order of trigger, which is that the trigger from the lower grid should occur before the anode trigger. 

\subsection{Fiducial Volume Cut}


The range of the alpha-particles in the counting gas P10 was  simulated using the SRIM ion range simulation package
\cite{SRIM}. The parameters that are used as input for the SRIM simulation are given in Table~\ref{tab:SRIM2}. 

\begin{table}
\begin{center}
\begin{tabular}{ll}  
   \toprule
  Parameter    & Value \\
   \midrule
   Composition [\%]      &         \\
   \hspace{0.5cm}Argon     &    90         \\
   \hspace{0.5cm}Hydrogen       & 8       \\
   \hspace{0.5cm}Carbon      &   2    \\
   Density [g/cm$^{3}$] & 0.00156          \\
   Number of events & 99,999 \\
 \end{tabular} 
 \end{center}
  \caption{Values used for SRIM simulation of P10. The majority of the spacial cuts were based on the resultant simulation. }
  \label{tab:SRIM2}
 \end{table}

\begin{figure}
\centering
\includegraphics[width=\linewidth]{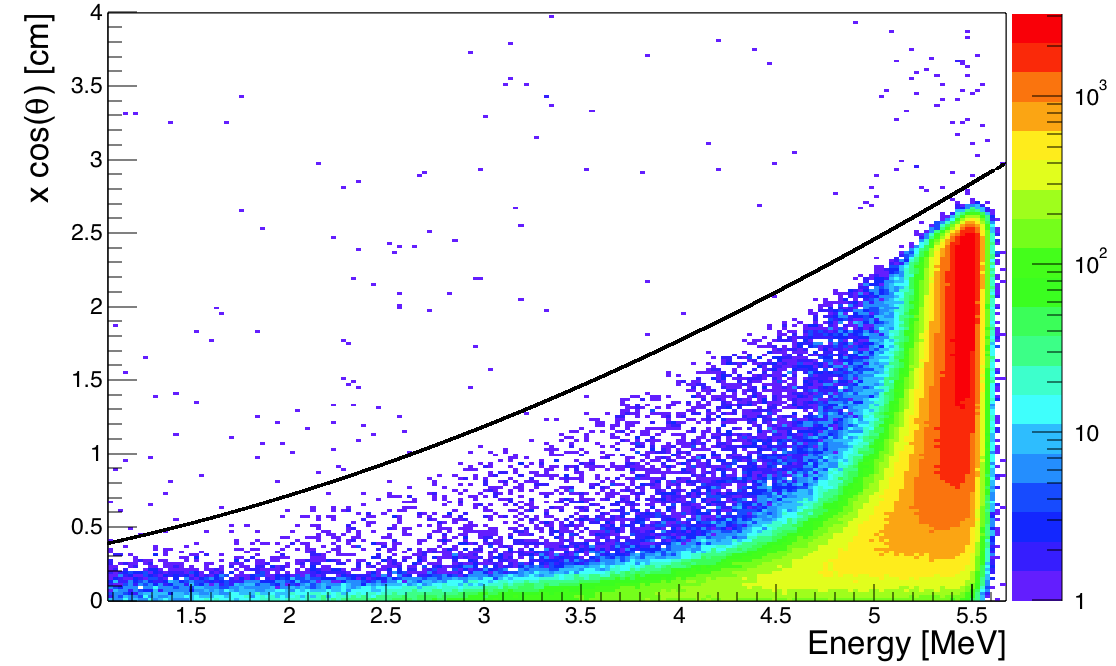}
\caption{Histogram showing the reconstructed position vs. energy. The reconstruction of the position follows equation~\ref{eq:pos}.The colour on the graph indicates the number of entries of each bin. The black line shows the centre of charge cloud simulated with SRIM. The events with a $\theta$ of zero are the closest to the black line. }
\label{fig:cut_pos}
\end{figure}

To determine the centre of the charge cloud, the mean was taken from the ionisation profile (dE/dx vs. range). This was done as most of the energy loss of the alpha particle is due to ionisation, so the centre of the charge cloud will be shifted towards the end point of the alpha track. This process was repeated for energies between 0.5~MeV and 9~MeV in steps of 0.5 MeV. The resulting values were fitted with a third order polynomial, and this was used as the basis of the fiducial volume cut. Figure~\ref{fig:cut_pos} shows the relation between the energy of the event and its relative position as well as the maximum of the fiducial volume cut. Due to the the smearing of $x \cos(\theta)$ related to the energy resolution, the minimum position value is set to below zero (-4~mm). 

\subsection{Drift Mobility Cut}\label{sec:drift}

The drift mobility of the electrons caused by the ionisation of the alpha particle in P10 can be calculated. This is done by using the value of the position, as well as the time between the grid and the anode trigger. The only change is that instead of the centre of the ionisation, the maximum range was used. This is because the grid is triggered when the electrons start to move, and the anode is triggered when the electrons from the ionisation cloud start to move inside the detection region. This means the anode is triggered by the electrons from the maximum range of the alpha path, and not the centre of the ionisation cloud. The difference between the centre of the ionisation cloud and its maximum size was also simulated through the use of SRIM. The electron drift mobility is given by Equation~\ref{eq:mu}. 

	\begin{equation}\label{eq:mu}
		\mu_{e} = \frac{\left( X_{0}- a(E)\cdot x\cdot\cos(\theta) \right)}{\left(T_{Anode}-T_{Grid}\right)} \;\;,
	\end{equation}

where $a(E)$ is the simulated correction function to change the centre of charge into the total range. 

The electron drift mobility in P10 is well documented and is dependent on E/p \cite{P10}, this is used to apply a cut to the data. The documented value for the electron drift mobility at ($0.105\pm0.004$) V/(cm$\cdot$Torr) is ($4.773 \pm 0.095$)~cm/$\mu$s in pure P10, the measured value is ($5.260\pm0.074$)~cm/$\mu$s. The uncertainty on the given value come from the possible ranges in pressure and voltage over a typical run period. The discrepancy between the two values could be due to the influence of contaminants in the gas. The P10 currently used is of a much purer standard than that which was typically available in the past. 
 
\begin{figure}
\centering
\includegraphics[width=\linewidth]{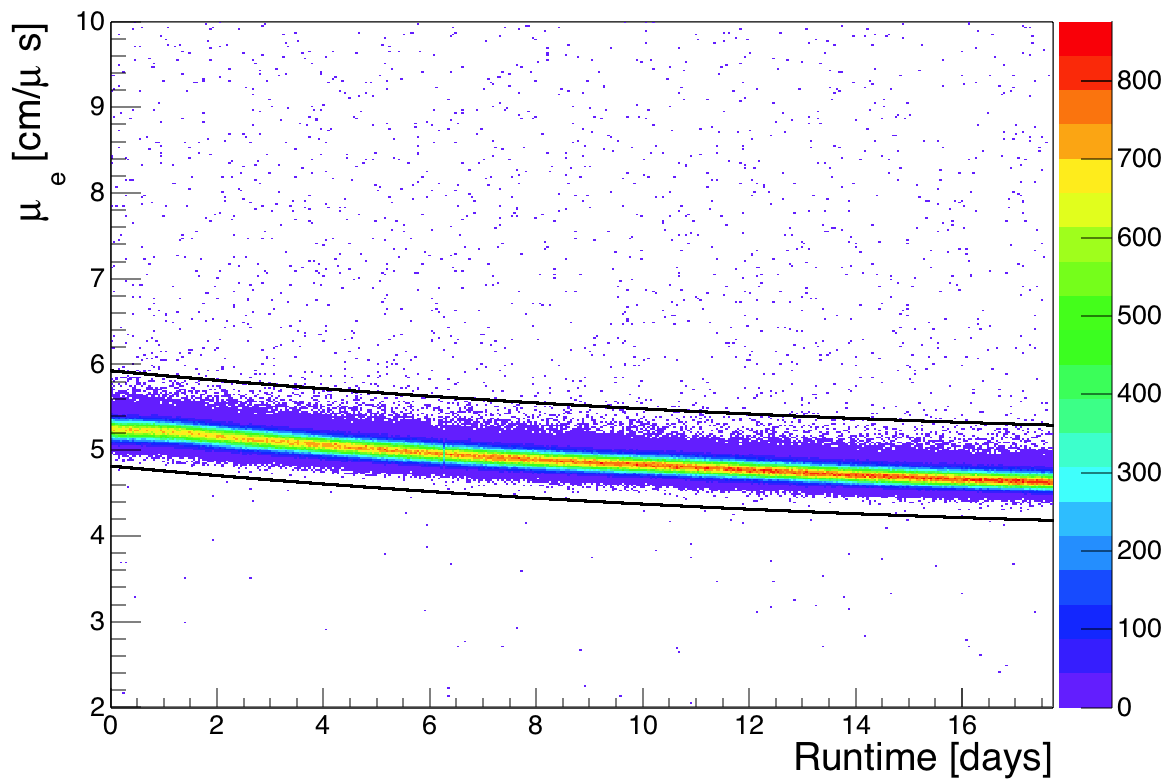}
\caption{Electron mobility $\mu_{e}$ as function of measuring time. The value of $\mu_{e}$ is determined for each event using equation~\ref{eq:mu}. The upper and lower values for the cut are shown with the two black lines. The cuts have been optimised using pulse shape analysis. All the events that fall above the upper cut originate from the walls of the chamber. The events that fall below the lower cut are due to mis-triggering pulses cause by micro-phonics.}
\label{fig:mu}
\end{figure}

The cut on the electron drift mobility is sensitive to the horizontal planar location of the alpha event. The walls of the chamber are at ground potential, this means that when an event occurs on the wall, the drifted electrons will experience a stronger electric field. This is because the distance between the wall of the chamber and the anode is less than the distance from the sample to the anode. This artificially gives a higher drift mobility ($\mu_{e}>6$ cm/$\mu$s). A test was performed with the $^{241}$Am source located as close to the chamber wall as possible. This showed a significant shift of the measured values of $\mu_{e}$ towards larger mobilities. A graph showing the electron drift mobility and the used cut is shown in Figure~\ref{fig:mu}.

There is a change in the value of $\mu_{e}$ as a function of measuring time, this is due to a small leak into the chamber
which was mentioned before. The effect can be described by Blanc's Law \cite{blacn}. Equation~\ref{eq:blanc} is used to determine the change in $\mu_{e}$ as a function of time.

\begin{equation}\label{eq:blanc}
		\mu_e(t) = \left[  \sum \frac{1}{\mu_i}\right]^{-1} \sim \frac{\mu_{\text{P10}}}{1+ p_1\cdot t + p_2 \cdot t^2} \; ,
\end{equation}

where $\mu_{\text{P10}}$ is the electron drift mobility in P10, and $t$ is the run time. The parameters $p_1$ and $p_2$ are obtained from fitting the $^{241}$Am calibration data set. The minimum and maximum cut for the $\mu_e$ is obtained by shifting the value up and down by a constant. The values of the two constants were optimised using the technique described in the next section.
\subsection{Cut Efficiency}
The cut efficiency is determined as follows. Several data sets from different runs, and from different times within the runs are selected at random. For each chosen run the pulse shapes are inspected by eye. The pulses are classified into different categories due to their causes. Their causes are signal, ``remaining'' and noise. Signal pulses fulfil the criteria of being caused by alpha events in the lower chamber. 

A ``remaining'' pulse is a pulse that is caused by a physical event inside of the chamber. This type of pulse does not meet the criteria to be a signal pulse. These pulses are usually distorted by the position of the event, or could be secondary decays caused by cosmic ray muons. 

Noise pulses are identified by being random oscillations on all of the FADC channel readouts. Due to the random nature of the noise, it is possible for a noise pulse to trigger correctly and survive the basic cuts. This fact emphasises the need for well developed cuts. 

Once these collections of pulses have been created, the cuts described in Sections \ref{sec:noi} to \ref{sec:drift} 
were applied to the pulses. The cut efficiency $\eta_{\text{cut}}$ is then determined using Equation~\ref{eq:eff}.

\begin{equation}\label{eq:eff}
	\eta_{\text{cut}} = \eta_{\text{signal}}\cdot \left( 1 - \eta_{\text{rem}} \right) \cdot \left( 1 - \eta_{\text{noise}} \right) \;,
\end{equation}

where $\eta_{signal}$ is the number of signal events that survive the cuts, $\eta_{\text{rem}}$ and $\eta_{\text{noise}}$ are the number of ``remaining" and noise events that are not removed by the cuts. A total of 6066 signal pulses, 1383 other pulses and 419 noise pulses were used to determine a total cut efficiency of 99.3~\%. 

The total efficiency for the chamber is then determined by comparing the activity of the source with the rate measured for the $^{241}$Am source. This is shown in Equation~\ref{eq:efftot}.

\begin{align}\label{eq:efftot}
	\eta_{\text{detection}} &= \frac{A_{\text{measured}}}{A_{\text{source}}} = (49.6 \pm 1.1) \;\; ,\%\\
	\eta_{\text{total}}&= \eta_{\text{detection}}   \cdot \eta_{\text{cut}}  = (49.3 \pm 1.1) \% \;\;. 
\end{align}

The main reduction in the efficiency is that there is only 2$\pi$ coverage of the sample. The efficiency is relatively high compared to other detection methods. The main source of uncertainty comes from the known precision of the $^{241}$Am activity. The efficiency uncertainty could be improved with the use of a better characterised alpha source. 

\section{Background Measurements}
\label{sec:back}

A background measurement was taken, the chamber was operated in the usual manor with no source present. The run showed no instability.  

\begin{figure}
\centering
\includegraphics[width=\linewidth]{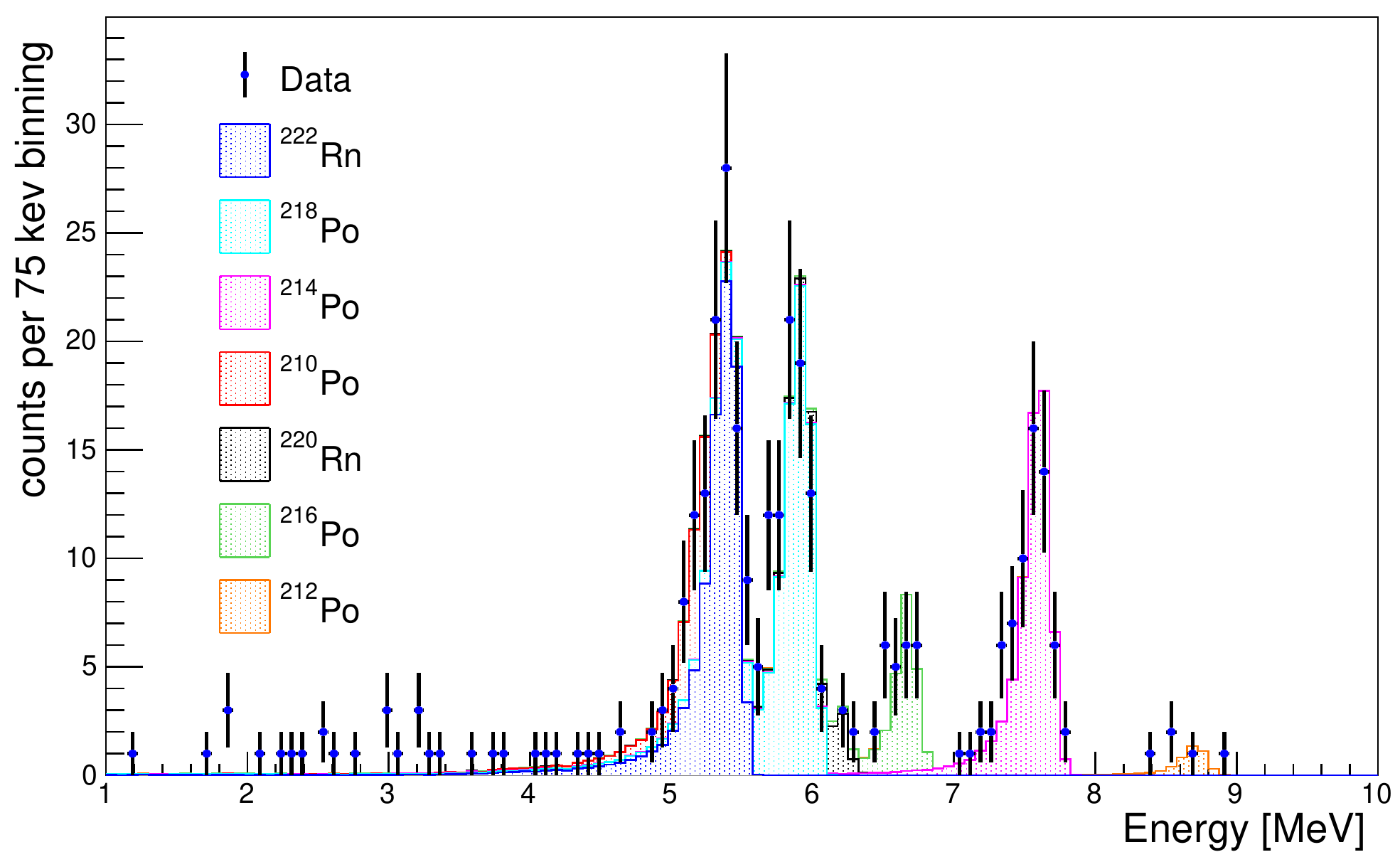}
\caption{Background spectrum from a 30.8~day run period. Fitted with background Monte Carlo spectra considering decay lines from the $^{238}$U and $^{232}$Th decay series. The values of the rates obtained from the fit are shown in Table~\ref{tab:bg}.}
\label{fig:background}
\end{figure}

\begin{table}[htbp]
\begin{center}
\begin{tabular}{l|c}
\hline
   \toprule
  Element    & Rate [c.p.d.] \\
   \midrule
$^{222}$Rn & 2.89  $\pm$  0.41 \\ 
$^{218}$Po & 2.77  $\pm$  0.32 \\ 
$^{214}$Po & 2.12  $\pm$  0.27 \\ 
$^{210}$Po & 0.85  $\pm$  0.31 \\ 
$^{220}$Rn & 0.27  $\pm$  0.16 \\ 
$^{216}$Po & 0.79  $\pm$  0.18 \\ 
$^{212}$Po & 0.16  $\pm$  0.11 \\ 
\end{tabular}
\caption{Values from the fit of the background Monte Carlo spectra data shown in Figure~\ref{fig:background}. The main contribution is from the daughters of $^{222}$Rn. There are also small contributions from $^{220}$Rn, coming from the $^{232}$Th decay chain.}
\label{tab:bg}
\end{center}
\end{table}

The obtained background spectrum is shown in Figure~\ref{fig:background}, it shows a total count number of 337 over the period of 30.8~days, this gives a background rate of ($10.9 \pm 0.6$) c.p.d. for the energy region of 1~MeV to 9~MeV, the error here is purely statistical. 

A background Monte Carlo was developed using the GEANT4 \cite{GEANT4} simulation package. This was used to simulate the energy spectra expected from different elements. The advantage of the simulation is an accurate determination of the energy tailing. It was assumed that all of the peaks had tailing caused by the same absorption thickness. The energy spectra were then smeared out with the detector resolution function. The data was fitted with the normalised simulated spectra, leaving only the overall area as a free parameter. The $\chi^2$/d.o.f for this fit is 48/56 = 0.86. The result of this fit with simulation is given in table \ref{tab:bg}.

The main contributions for the background come from $^{222}$Rn daughters ($^{218}$Po, $^{214}$Po and $^{210}$Po). The radon is present in the lab environment in gaseous form entering through the before mentioned leak. In addition, when the samples are loaded some of the gas stays in the detector. Repeated flushing also helps to diminish this background contribution. There are also contributions from the $^{232}$Th decay chain ($^{220}$Rn, $^{216}$Po and $^{212}$Po). This decay is thought to originate from dust particles in the laboratory environment, but could also be coming from the holder. The contribution form $^{232}$Th is a small part of the total background rate.   

The background rate can further be reduced by removing the first 12~days and only using the last 18.8~days. The background activity falls to ($8.3\pm0.7$) c.p.d. in the same energy region. This is due to the radon daughters decaying to longer living elements. 
 
\section{Summary and Conclusions}
\label{sec:sum}

The above work shows the successful construction of a low-radiation chamber specifically designed to measure long living alpha-decays. The sensitivity to these decays have been enhanced with the use of data analysis cuts based on the physical parameters of the chamber and the properties of the P10. The cuts are based on the drift mobility of the ionised electrons in the P10, as well as the position and angle of the event. 

Investigations are ongoing into the cause of the leak into the chamber. Future work will be done to run the chamber in constant flush mode. For current measurements the stability of the chamber is shown to be adequate. 

\section{Acknowledgement} 
This work was supported by BMBF 02NUK13A HZDR and 02NUK13B TUD.
We thank L. Heinrich and M. Paul at Helmholtz-Zentrum Dresden-Rossendorf (HZDR) for the help with the design and construction of the chamber as well as A. Junghans for valuable discussions. 
We also thank the PTB-Braunschweig and Dr. Taut (TU-Dresden) for providing us with $^{241}$Am sources.  

\bibliographystyle{elsarticle-num} 
\bibliography{Chamber.bib}


%
%
%
\end{document}